\documentclass[twocolumn,aps,prd,10pt,superscriptaddress,longbibliography]{revtex4-1}

\usepackage{amssymb,amsmath}
\usepackage{bm}
\usepackage{dcolumn}
\usepackage{glossaries}
\usepackage{graphicx}
\usepackage[version=4]{mhchem}
\usepackage[usenames,dvipsnames,svgnames]{xcolor}
\usepackage{siunitx}
\usepackage{hyperref}
\usepackage{makecell}
\hypersetup{
pdfnewwindow=true,      
colorlinks=true,        
linkcolor=Blue,         
citecolor=Blue,         
filecolor=Blue,         
urlcolor=Blue           
}

\newacronym{mof}{MOF}{metal-organic framework}
\newacronym{mfp}{MFP}{mean free path}
\newacronym{sed}{SED}{spectral energy density}
\newacronym{dft}{DFT}{density functional theory}
\newacronym{nep}{NEP}{neuroevolution potential}
\newacronym{bte}{BTE}{Boltzmann transport equation}
\newacronym{ald}{ALD}{anharmonic lattice dynamics}
\newacronym{sm}{SM}{Supplemental Material}
\newacronym{md}{MD}{molecular dynamics}
\newacronym{mlp}{MLP}{machine-learned potential}
\newacronym{2d}{2D}{two-dimensional}
\newacronym{hnemd}{HNEMD}{homogeneous non-equilibrium molecular dynamics}
\newacronym{nemd}{NEMD}{non-equilibrium molecular dynamics}
\newacronym{rmse}{RMSE}{root mean square error}
\newacronym{ltc}{LTC}{lattice thermal conductivity}
\newacronym{vdos}{VDOS}{vibrational density of states}

\DeclareSIUnit\angstrom{\text{Å}}
\DeclareSIUnit{\atom}{atom}
\DeclareSIUnit{\step}{step}
\DeclareSIUnit{\atomstepsecond}{\atom\step\per\second}

\begin{document}

\title{Structurally Triggered Breakdown of the Phonon Gas Model in Crystalline Metal-Organic Frameworks}

\author{Penghua Ying}
\email{penghua@xjtu.edu.cn}
\affiliation{Laboratory for multiscale mechanics and medical science, SV LAB, School of Aerospace, Xi’an Jiaotong University, Xi’an, Shaanxi, 710049, China}

\author{Ting Liang}
\affiliation{Department of Electronic Engineering and Materials Science and Technology Research Center, The Chinese University of Hong Kong, Shatin, N.T., Hong Kong SAR, 999077, P. R. China}


\author{Yun Chen}
\email{yunchen125@gmail.com}
\affiliation{Department of Physical Chemistry, School of Chemistry, Tel Aviv University, Tel Aviv, 6997801, Israel}

\author{Yan Chen}
\affiliation{Laboratory for multiscale mechanics and medical science, SV LAB, School of Aerospace, Xi’an Jiaotong University, Xi’an, Shaanxi, 710049, China}

\author{Shiyun Xiong}
\affiliation{Guangzhou Key Laboratory of Low-Dimensional Materials and Energy Storage Devices, School of Materials and Energy, Guangdong University of Technology, Guangzhou 510006, China}

\author{Zheyong Fan}
\affiliation{College of Physical Science and Technology, Bohai University, Jinzhou 121013, P. R. China}

\author{Jianbin Xu}
\affiliation{Department of Electronic Engineering and Materials Science and Technology Research Center, The Chinese University of Hong Kong, Shatin, N.T., Hong Kong SAR, 999077, P. R. China}

\author{Yilun Liu}
\email{yilunliu@mail.xjtu.edu.cn}
\affiliation{Laboratory for multiscale mechanics and medical science, SV LAB, School of Aerospace, Xi’an Jiaotong University, Xi’an, Shaanxi, 710049, China}

\date{\today}

\begin{abstract}

While crystalline materials with glass-like thermal conductivity are fundamentally intriguing, structurally triggering the transition from propagating to diffusive heat transport within a single framework remains a formidable challenge. Here, using extensive machine learning molecular dynamics, we demonstrate a fundamental thermal transport crossover in metal-organic frameworks. We reveal that grafting flexible side chains onto a pristine MOF backbone acts as a structural switch, strongly reducing the thermal conductivity by $\sim$70\% (from $\sim 0.7$ to $\sim 0.2\ \text{W m}^{-1}\text{K}^{-1}$ at \SI{300}{\kelvin}). 
Crucially, the functionalized derivatives exhibit a drastic transition from a classical Peierls $\sim 1/T$ decay to an anomalous, temperature-independent glass-like plateau. Reciprocal- and real-space analyses reveal the microscopic origins: the side chains act as built-in local resonators that trap acoustic energy via strong low-frequency resonant hybridization, while simultaneously inducing extreme steric crowding. Consequently, the heat-carrying phonon modes become critically damped, with their mean free paths strictly confined to the nanometer scale and their lifetimes collapsing to the Ioffe-Regel limit. This work establishes a highly programmable molecular engineering strategy to dismantle the phonon gas model, forcing crystalline frameworks into an extreme diffusive transport regime.

\end{abstract}
\maketitle
\raggedbottom

\section{Introduction}

\Glspl{mof} represent a unique class of ``designer solids'' where rigid inorganic nodes are bridged by highly tunable organic linkers, offering unprecedented bottom-up control over the crystalline lattice~\cite{yaghi2003reticular}. This extreme structural programmability, combined with their inherently low thermal conductivity, positions them as promising candidates for advanced thermal management and thermoelectric applications~\cite{erickson2015thin, fan2021recent,erickson2015thin}. Despite their complex unit cells and strong intrinsic anharmonicity typically yielding extreme thermal resistance, recent investigations present a counterintuitive transport picture: phonon \glspl{mfp} in \gls{mof} crystals can surprisingly extend to the sub-micrometer scale~\cite{fan2022ultralong, ying2023sub}. The persistence of these long-range phonon propagation channels indicates that the pristine topological skeleton alone is insufficient to reach the fundamental minimum of thermal conductivity. While structural amorphization provides a trivial route to minimize heat transfer, achieving glass-like thermal transport while preserving long-range crystalline order remains a formidable challenge. Consequently, effectively truncating these sub-micrometer transport channels through precise molecular engineering, such as the functionalization of ligands with organic side chains~\cite{pallach2021frustrated}, to push the thermal conductivity toward its amorphous limit without sacrificing crystalline integrity has become a critical pursuit in contemporary phonon engineering.

Traditionally, thermal transport in dielectric crystals is described by the Peierls-\gls{bte}~\cite{mcgaughey2019phonon, folkner2024elastic, zeng2025thermal}. However, for systems with massive primitive cells or intense structural disorder (e.g., frustrated flexibility~\cite{pallach2021frustrated}), this classical quasiparticle picture breaks down~\cite{simoncelli2019unified, lucente2023crossover}. When severe scattering reduces the mean free path to the interatomic spacing (the Ioffe-Regel limit)~\cite{ioffe1960non, luckyanova2018phonon}, and extreme spectral broadening induces band overlap, the wave-like nature of phonons dominates. The unified Wigner formulation~\cite{simoncelli2022wigner} provides the necessary framework to capture the coexistence of particle-like propagation and wave-like coherent tunneling. Yet, to computationally capture this particle-wave crossover, perturbative \gls{ald} approaches become prohibitively expensive for \glspl{mof}, as their exceptionally large primitive cells make high-order force-constant extraction intractable. While \gls{md} inherently incorporates arbitrary-order scattering and coherent transport~\cite{gu2021thermal,zhang2022heat}, it has traditionally faced a trade-off between accuracy and computational efficiency. \Glspl{mlp}~\cite{behler2016perspective, friederich2021machine} resolve this bottleneck by achieving near-\gls{dft} accuracy with empirical efficiency, making the direct, large-scale \gls{md} simulation of extreme anharmonicity and coherent transport computationally feasible~\cite{dong2024molecular}. In this Letter, utilizing \gls{mlp}-driven \gls{md} simulations, we reveal a fundamental crossover from crystalline to glass-like thermal transport in \glspl{mof} induced by side-chain functionalization.



\begin{figure}[htb]
\begin{center}
\includegraphics[width=\columnwidth]{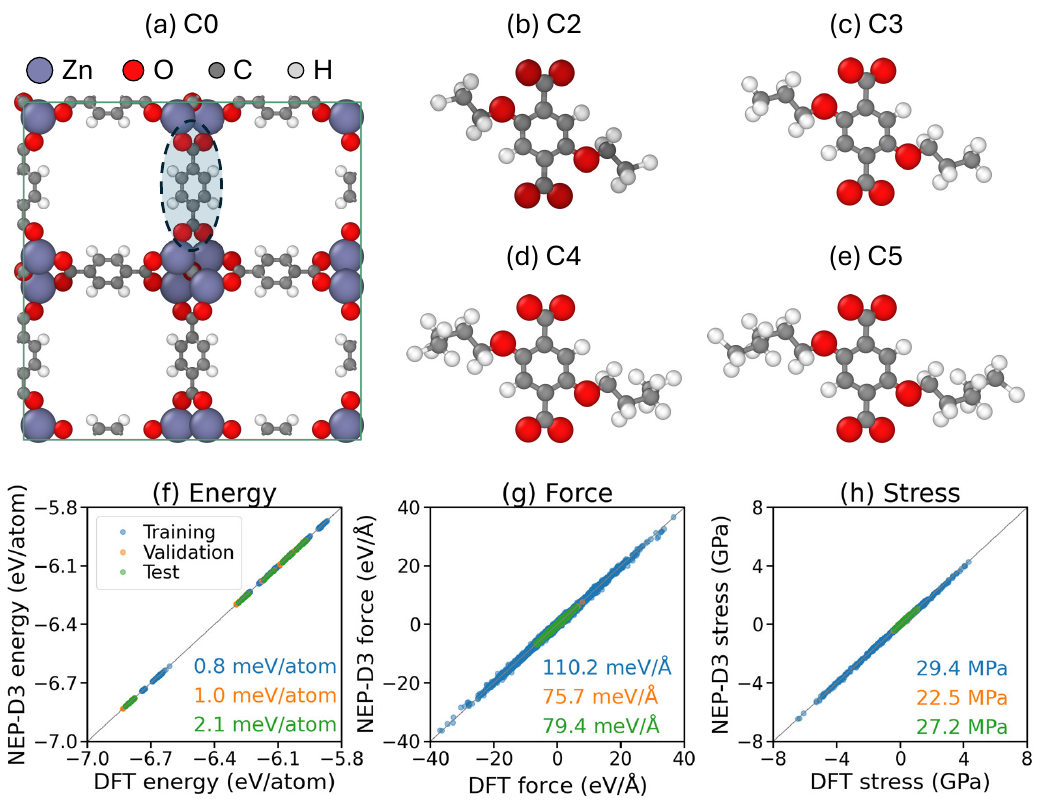}
\caption{Structural models and machine-learning potential validation. (a)--(e) Atomistic structures of (a) the pristine MOF-5 framework (C0) and its functionalized derivatives incorporating alkoxy side chains with (b) two (C2), (c) three (C3), (d) four (C4), and (e) five (C5) carbon atoms. The OVITO package~\cite{stukowski2010visualization} was used for visualization. (f)--(h) Parity plots comparing the NEP+D3 predictions against \gls{dft} reference calculations for (f) energies, (g) atomic forces, and (h) stresses across the training, validation, and test datasets.}
\label{fig:model}
\end{center}
\end{figure}

To investigate this transport crossover, we focus on pristine MOF-5 (the archetypal \gls{mof}~\cite{li1999design}, denoted as C0 here) and a series of its experimentally synthesized derivatives~\cite{pallach2021frustrated}. The pristine C0 features a cubic topology, where massive inorganic $\mathrm{Zn_{4}O}$ clusters act as rigid nodes bridged by 1,4-benzenedicarboxylate organic linkers. Building upon this foundational architecture, the functionalized derivatives feature linear alkoxy side chains of varying lengths---ethoxy (C2), propoxy (C3), butoxy (C4), and pentoxy (C5)---covalently grafted onto the organic linkers (\autoref{fig:model}(a)--(e)). To accurately capture the strong anharmonicity of these systems, we developed a unified machine-learning potential utilizing the \gls{nep} framework~\cite{fan2021neuroevolution,fan2022improving} augmented with the D3 dispersion correction~\cite{grimme2010jcp,grimme2011effect} (\gls{nep}+D3~\cite{ying2024combining}). By explicitly partitioning the atomic interactions into a short-range neural-network representation and a long-range analytical van der Waals correction, the \gls{nep}+D3 model ensures superior simulation robustness and predictive accuracy~\cite{ying2024combining}. The \gls{nep} framework was specifically selected for its exceptional computational efficiency, an essential prerequisite for the extensive \gls{md} sampling required in thermal transport simulations~\cite{wang2026interfacial, jiang2026chirality}. We used an active learning strategy on \gls{dft}-labeled structures to thoroughly sample the complex configurational spaces of C0 through C5  (see \gls{sm} Note S1 for details). Our model achieves near-\gls{dft} accuracy, keeping \gls{rmse} for energies, atomic forces, and stresses across training, validation, and test datasets below \SI{2.2}{\milli\electronvolt\per\atom}, \SI{111}{\milli\electronvolt\per\angstrom}, and \SI{30}{\mega\pascal}, respectively (\autoref{fig:model}(f)--(h)). All \gls{nep}+D3 model training and subsequent \gls{md} simulations were conducted using the \textsc{gpumd} package~\cite{fan2022gpumd,xu2025gpumd}.


\begin{figure}[htb]
\begin{center}
\includegraphics[width=\columnwidth]{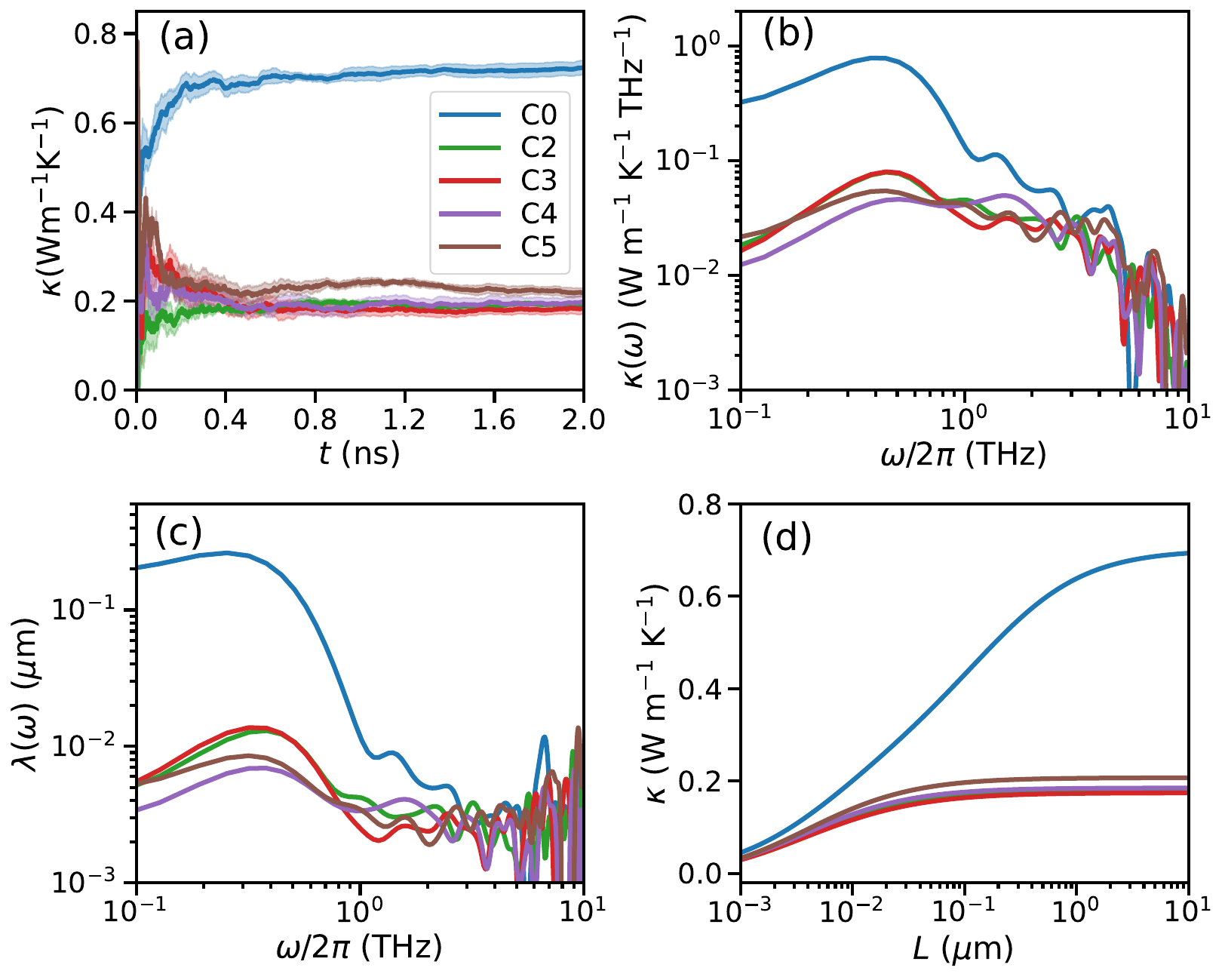}
\caption{Thermal transport at \SI{300}{\kelvin}. (a) Running thermal conductivity as a function of simulation time. (b) Spectral thermal conductivity and (c) derived phonon \gls{mfp} as a function of phonon frequency. (d) Apparent thermal conductivity as a function of the transport length, illustrating the ballistic-to-diffusive crossover. Panels (a) and (b) are calculated using \gls{hnemd} simulations, while panels (c) and (d) are derived from both \gls{hnemd} and \gls{nemd} simulations. Solid lines denote the average from three independent simulations, with the shaded area in panel (a) representing the standard error of the mean.}
\label{fig:ltc}
\end{center}
\end{figure}

To evaluate the thermal transport properties of \glspl{mof}, we employ the \gls{hnemd}~\cite{fan2019homogeneousa} and \gls{nemd} methods (see \gls{sm} Notes S2 and S3 for comprehensive theoretical formalisms and simulation protocols). While formally equivalent to the classic Green-Kubo formalism~\cite{green1954markoff,kubo1957statistical}, \gls{hnemd} offers significantly superior computational efficiency and statistical convergence. Because eliminating finite-size effects is critical for accurate thermal conductivity ($\kappa$) prediction, we performed rigorous size-convergence tests (\gls{sm} Figure S1). We found that a large $7\times 7\times 7$ supercell (\num{171157} atoms) is required to capture the long-range phonon channels in pristine C0, whereas $4\times 4\times 4$ supercells are sufficient for the functionalized \glspl{mof}. At \SI{300}{\kelvin}, the pristine C0 yields a fully converged $\kappa$ of \SI{0.72 \pm 0.02}{{W m^{-1} K^{-1}}} (\autoref{fig:ltc}(a)). This value exceeds the previously reported one \SI{0.62\pm 0.01}{{W m^{-1} K^{-1}}}~\cite{ying2023sub}, which was constrained by a smaller $5\times 5\times 5$ simulation domain. Notably, the overall thermal transport is dominated by the short-range \gls{nep} interactions. While the D3 dispersion correction reduces the $\kappa$ of pristine C0 by $\sim$10\% (consistent with Ref.~\citenum{ying2024combining}), it has a minimal effect on the functionalized derivatives (see Figure S2 in the \gls{sm}).

Strikingly, the grafting of alkoxy side chains induces a precipitous drop in \gls{ltc}. The $\kappa$ of the functionalized derivatives (C2 through C5) are suppressed by $\sim$70\%, tightly clustering around an ultralow value of $\sim$\SI{0.20}{{W m^{-1} K^{-1}}} (\autoref{fig:ltc}(a)). To unravel the microscopic origins of this drastic thermal suppression, we spectrally decompose~\cite{fan2019homogeneousa} the \gls{hnemd} heat current to map the frequency contributions (\autoref{fig:ltc}(b)), revealing that the $\kappa$ for all \glspl{mof} is dominated by phonon frequencies below \SI{10}{\tera\hertz}. By combining this with the ballistic spectral thermal conductance (see \gls{sm} Figure S3) derived from \gls{nemd}, we quantitatively determine the distributions of phonon \glspl{mfp} (\autoref{fig:ltc}(c)) and the corresponding length-dependent thermal conductivity (see \gls{sm} Note S3 for details). For pristine C0, low-frequency phonons with \glspl{mfp} exceeding \SI{100}{\nano\meter} carry most of the heat, confirming robust sub-micrometer propagation channels~\cite{ying2023sub}. However, side chains abruptly truncate these long-range channels. The maximum \glspl{mfp} in C2 through C5 are reduced by over an order of magnitude, down to within $\sim$\SI{10}{\nano\meter}. Consequently, as illustrated in \autoref{fig:ltc}(d), while the apparent thermal conductivity of C0 exhibits a prolonged length dependence characteristic of a broad ballistic-to-diffusive crossover at the micrometer scale, the transport in the functionalized \glspl{mof} reaches the diffusive limit rapidly at the nanometer scale. Their apparent thermal conductivities converge at remarkably short lengths, a hallmark of extreme acoustic damping.

\begin{figure}[htb]
\begin{center}
\includegraphics[width=1\columnwidth]{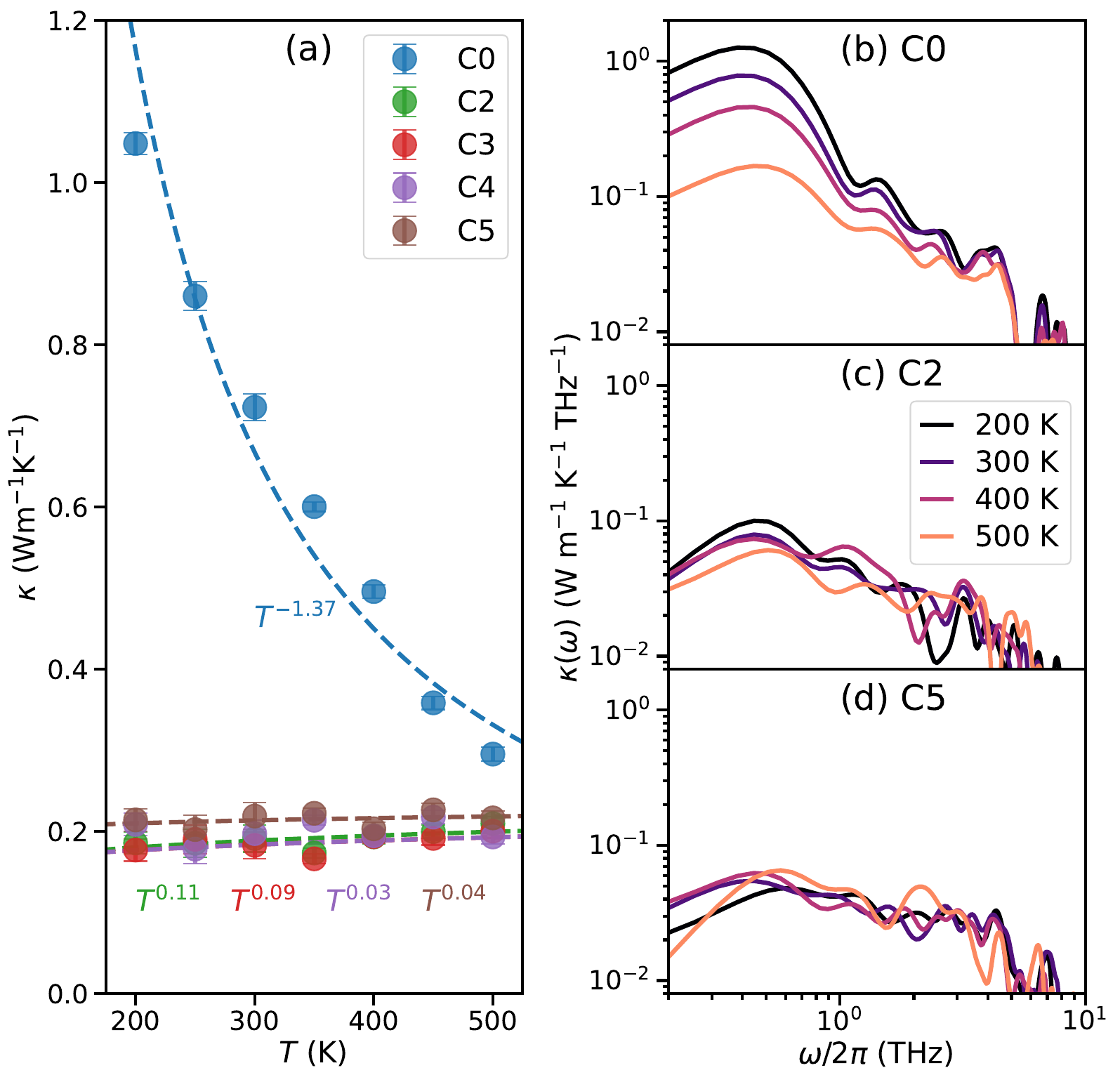}
\caption{Transition from temperature-dependent to temperature-independent thermal conductivity. (a) Thermal conductivity of MOF-5 structures with varying side-chain lengths as a function of temperature, with dashed lines indicating power-law fits. (b)--(d) Spectral thermal conductivity as a function of phonon frequency for (b) pristine C0, (c) C2, and (d) C5 at four representative temperatures: \SI{200}{\kelvin}, \SI{300}{\kelvin}, \SI{400}{\kelvin}, and \SI{500}{\kelvin}.}
\label{fig:temp}
\end{center}
\end{figure}

The most compelling evidence of the particle-wave transition is revealed in the temperature dependence of the \gls{ltc} (\autoref{fig:temp}(a)). For pristine C0, $\kappa$ follows a typical power-law decay of $T^{-1.37}$, a characteristic signature of crystalline transport dominated by anharmonic Umklapp scattering. In stark contrast, the functionalized systems (C2--C5) show an anomalous, glass-like behavior: their \gls{ltc} remain almost constant from \SI{200}{\kelvin} to \SI{500}{\kelvin}. Their fitted temperature exponents are nearly zero ($T^{0.03}$ to $T^{0.11}$), indicating that the classical quasiparticle scattering mechanism has reached its saturation limit. 
While temperature-independent, glass-like thermal transport is known in specific systems including disordered solids~\cite{simonov2020designing, liang2023prb}, complex crystals like $\mathrm{Tl_3VSe_4}$~\cite{mukhopadhyay2018two, simoncelli2019unified}, and superionic materials~\cite{zhou2025heat}, triggering this extreme transition within a single crystalline framework simply by grafting side chains highlights the unique structural programmability of \glspl{mof}.

The spectral thermal conductivity, $\kappa(\omega)$, explains this macroscopic transition (\autoref{fig:temp}(b)--(d)). While the spectral peaks of C0 systematically drop as temperature increases, the functionalized systems behave very differently. In C2, the low-frequency peaks still decrease slightly with heating. This marks the last trace of particle-like transport before complete saturation. In contrast, C5 shows a complete crossover: at low frequencies, its spectral conductivity at \SI{500}{\kelvin} actually exceeds the values at lower temperatures. This unusual ``positive'' temperature dependence suggests that stronger anharmonicity at high temperatures broadens the vibrational modes. This broadening likely increases their spectral overlap, which would enhance the wave-like coherent tunneling channels. Ultimately, as the side chains grow longer, we hypothesize that particle-like channels completely saturate, forcing the system into a transport regime fully dominated by wave-like tunneling.

\begin{figure}[htb]
\begin{center}
\includegraphics[width=\columnwidth]{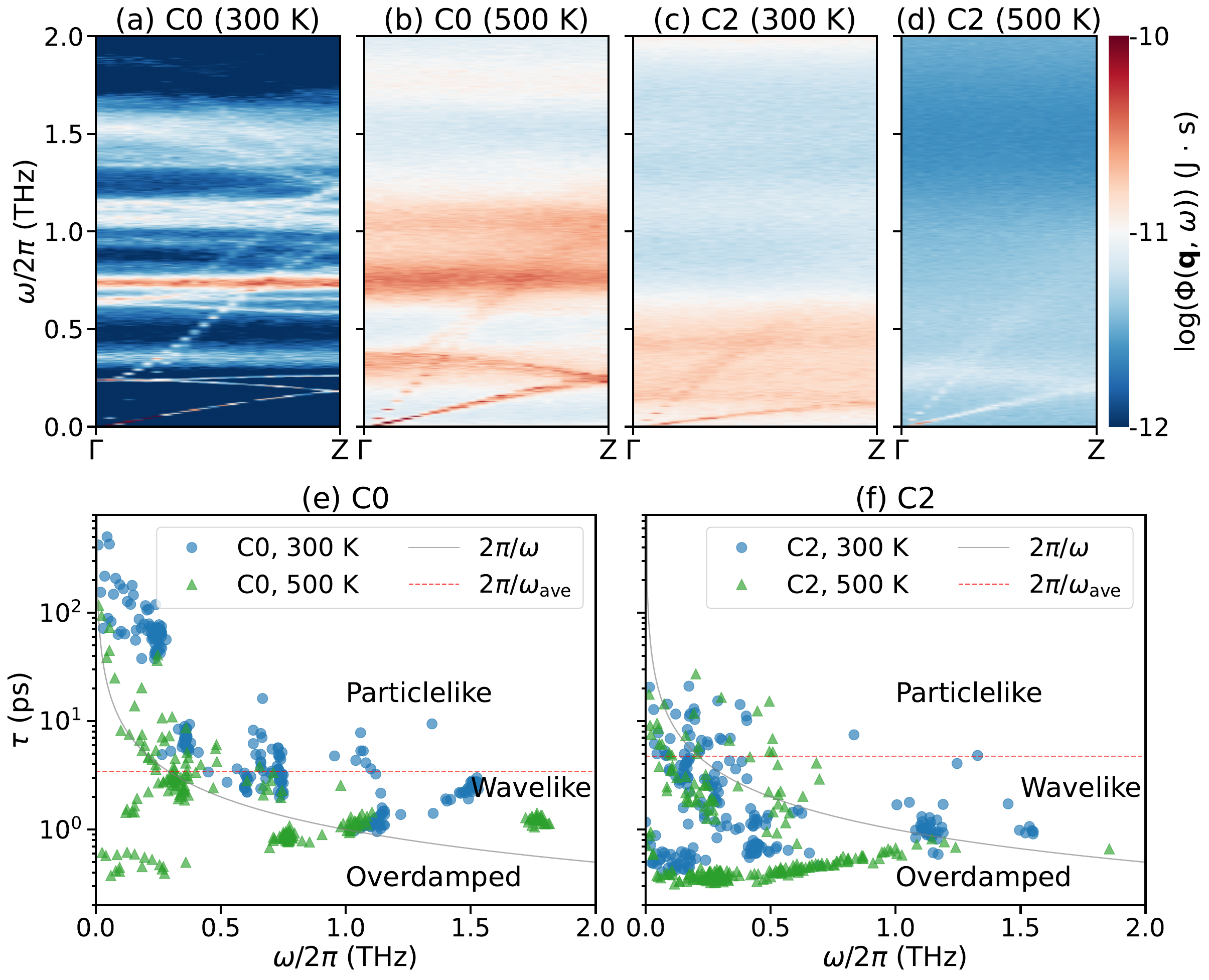}
\caption{\Gls{sed} and phonon lifetimes revealing the transition of thermal transport mechanisms. (a)--(d) Contour maps of the logarithm of \gls{sed} along the $\Gamma$-Z high-symmetry direction for (a) C0 at \SI{300}{\kelvin}, (b) C0 at \SI{500}{\kelvin}, (c) C2 at \SI{300}{\kelvin}, and (d) C2 at \SI{500}{\kelvin}. (e)–(f) Extracted phonon lifetimes ($\tau$) as a function of frequency for (e) C0 and (f) C2 at \SI{300}{\kelvin} (blue circles) and \SI{500}{\kelvin} (green triangles). The solid gray curve represents the Ioffe-Regel limit ($\tau = 1/\omega$), and the red dashed line denotes the Wigner limit ($\tau = 1/\omega_{\text{ave}}$)~\cite{simoncelli2022wigner}. These limits demarcate the distinct thermal transport regimes: particlelike propagation, wavelike tunneling, and overdamped modes.}
\label{fig:sed}
\end{center}
\end{figure}

To validate this hypothesis and uncover the microscopic transition mechanism, we used \gls{sed} analysis~\cite{thomas2010predicting} (via the \textsc{pysed} package~\cite{liang2025pysed}) to directly extract the phonon dispersions and mode-resolved lifetimes ($\tau$) for C0 and C2 from \gls{md} trajectories (\autoref{fig:sed}; see \gls{sm} Note S4 for details). In pristine C0 at \SI{300}{\kelvin}, the \gls{sed} contour maps exhibit sharp, well-defined dispersion branches (\autoref{fig:sed}(a)), consistent with Ref.~\citenum{ying2023sub}. Heating to \SI{500}{\kelvin} increases anharmonicity, which broadens these spectral peaks (\autoref{fig:sed}(b)) and drops the phonon lifetimes (\autoref{fig:sed}(e)). This directly explains the macroscopic  $T^{-1.37}$ thermal decay observed in \autoref{fig:temp}(a).

\begin{figure}[htb]
\begin{center}
\includegraphics[width=\columnwidth]{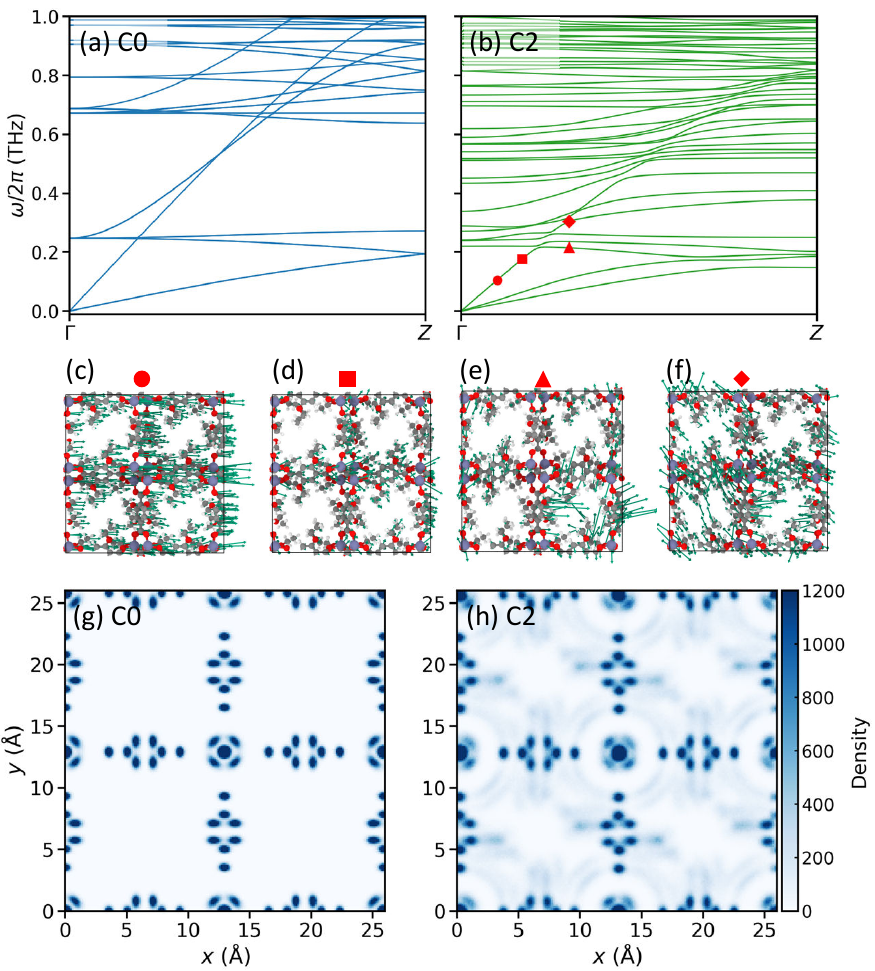}
\caption{Dual microscopic mechanisms driving the breakdown of the phonon gas model. (a,b) Phonon dispersion relations of the (a) pristine C0 and (b) functionalized C2 frameworks along the $\Gamma$-Z high-symmetry direction. (c)-(f) Visualizations of the vibrational eigenvectors for selected modes marked in (b). (g,h) Real-space \gls{2d} occurrence density maps of all carbon atomic trajectories folded into a single unit cell for (g) C0 and (h) C2 at \SI{300}{\kelvin}.}
\label{fig:real}
\end{center}
\end{figure}

However, adding flexible alkoxy side chains shatters this conventional temperature-dependent picture. In C2 at \SI{300}{\kelvin}, the \gls{sed} contours blur significantly (\autoref{fig:sed}(c)), revealing extreme acoustic damping. The dynamic rattling of the side chains suppresses the phonon lifetimes in C2 by orders of magnitude compared to C0. To quantitatively categorize this transport transition, we delineate the vibrational regimes using two fundamental theoretical bounds: the Wigner limit ($\tau = 1/\omega_{\text{ave}}$, determined by the average phonon frequency $\omega_{\text{ave}}$) which marks the onset of wave-like coherent tunneling, and the Ioffe-Regel limit ($\tau = 1/\omega$, representing extreme scattering within a single vibrational period) which defines the threshold for overdamped vibrations~\cite{simoncelli2022wigner}. This forces most vibrational modes in C2 below the Wigner limit into the wave-like tunneling regime, with a substantial fraction plunging even deeper below the Ioffe-Regel limit into the strictly overdamped regime (\autoref{fig:sed}(f)). Crucially, heating C2 to \SI{500}{\kelvin} smears the \gls{sed} map into a nearly featureless continuum (\autoref{fig:sed}(d)), yet the extracted lifetimes tightly overlap with those at \SI{300}{\kelvin} (\autoref{fig:sed}(f)). Numerous modes are pushed down to these theoretical minimums and remain trapped there regardless of the temperature. This temperature invariance of $\tau$ at the theoretical minimum provides our definitive microscopic evidence: the side-chain engineering saturates scattering and suppresses particle-like propagation channels.

To understand the origin of this lifetime collapse, we investigate both the reciprocal-space lattice dynamics (\gls{sm} Note S6) and the real-space atomic trajectories (\autoref{fig:real}; \gls{sm} Note S7). At \SI{0}{\kelvin}, the phonon dispersion of pristine C0 exhibits highly dispersive, continuous transverse and longitudinal acoustic (TA and LA) branches (\autoref{fig:real}(a)), characteristic of classic crystalline thermal transport. In stark contrast, C2 introduces dense, ultra-low-frequency flat optical bands at extremely low frequencies (below \SI{0.4}{\tera\hertz}) (\autoref{fig:real}(b)). Originating from the intrinsic vibrations of the alkoxy side chains, these flat bands strongly hybridize with the heat-carrying acoustic modes, creating prominent avoided crossings between \SI{0.2}{\tera\hertz} and \SI{0.3}{\tera\hertz}. Consequently, the phonon group velocities for the functionalized derivatives are drastically suppressed by nearly an order of magnitude (see \gls{sm} Figure S4). This reciprocal-space flattening directly manifests in the \gls{vdos} as a massive accumulation of ultra-low-frequency states. Unlike the pristine framework, the functionalized derivatives exhibit severe spectral broadening and a catastrophic deviation from the classical Debye scaling law ($g(\nu) \propto \nu^2$) in the low-frequency limit (see \gls{sm} Figure S5). This overpopulation of non-propagating, localized states perfectly corroborates the smeared \gls{sed} contours and the corresponding collapse of phonon lifetimes. 

Visualizations of the vibrational eigenvectors confirm this resonant energy capture. While typical acoustic modes drive collective framework translations (\autoref{fig:real}(c,d)), the hybridized modes at the avoided crossings (\autoref{fig:real}(e,f)) strictly localize vibrational energy on the side chains. Acting as built-in local resonators, these side chains trigger intense acoustic damping, a resonance mechanism fundamentally analogous to heat suppression in nanophononic metamaterials~\cite{davis2014nanophononic, xiong2016blocking}. This frequency-domain hybridization effectively localizes the acoustic energy, halting macroscopic phonon propagation.

We also project the finite-temperature (\SI{300}{\kelvin}) atomic dynamics into real space by mapping the \gls{2d} occurrence density of all carbon atoms (see \gls{sm} Note S7 for details). In pristine C0, the rigid framework maintains a clean, open pore architecture. The atoms vibrate tightly around their equilibrium nodes, preserving instantaneous translational symmetry (\autoref{fig:real}(g)). The functionalized C2 system, however, suffers from extreme spatial congestion (\autoref{fig:real}(h)). The highly flexible alkoxy side chains sweep through massive, uncoordinated configurations. Rather than remaining localized, their dynamic trajectories sprawl across and physically saturate the available free volume within the pores. This continuous, large-amplitude dynamic disorder (or steric crowding) shatters the periodic potential required for long-range phonon propagation.

Together, these dual mechanisms, namely frequency-domain resonant hybridization and real-space steric crowding, provide a vivid physical picture for the smeared \gls{sed} continuum and the breakdown of the phonon gas model. Following this C0-to-C2 trend, the extended side chains in C3 through C5 will clearly intensify this localized rattling effect. As corroborated by the 2D density maps (\gls{sm} Figure S6), the spatial delocalization becomes progressively more extreme, with the ultra-flexible pentoxy chains in C5 almost completely smearing across and filling the available pore volume. They will plunge an even greater fraction of vibrational modes deep into the overdamped regime. Ultimately, this structurally triggered dynamic disorder dismantles the classical transport framework, leaving temperature-independent wave-like and overdamped modes to drive the glass-like thermal transport.

\section{Summary and conclusions}
\label{section:summary}

In summary, we have unveiled the structurally triggered breakdown of the classical phonon gas model in \glspl{mof} through precise side-chain engineering. By bridging macroscopic thermal responses with reciprocal- and real-space microscopic dynamics, we demonstrate that the intense localized rattling of functionalized alkoxy chains induces extreme acoustic damping. This structural agitation paralyzes traditional quasiparticle channels at the Ioffe-Regel limit, forcing a fundamental crossover to a glass-like transport regime governed entirely by temperature-independent coherent and overdamped modes.

Given their vast chemical versatility, \glspl{mof} serve as an ideal platform for this targeted molecular manipulation. By leveraging localized dynamics to saturate phonon scattering, this structural design strategy drives crystalline thermal conductivity to its fundamental amorphous limit without sacrificing long-range topological order. Crucially, unlike traditional approaches such as defect engineering or amorphization that inevitably disrupt the global lattice, this metamaterial-like design effectively decouples the static lattice structure from dynamic thermal transport. By relying strictly on the localized rattling of side chains to suppress heat flow, the intrinsic topological integrity of the crystalline \gls{mof} backbone is remarkably preserved. Ultimately, this approach establishes a highly programmable design space for next-generation ultralow thermal conductivity materials, with profound implications for advanced thermal management and thermoelectrics.

\begin{acknowledgments}
This work was supported by the financial support of National Natural Science Foundation of China (12325204).
T. Liang and J. Xu acknowledge support from the National Key R\&D Project from the Ministry of Science and Technology of China (No. 2022YFA1203100), the Research Grants Council of Hong Kong (No. AoE/P-701/20), RGC GRF (No. 14220022, JLFS/E-402/24), and the CUHK Postdoctoral Fellowship.
\end{acknowledgments}

\noindent{\textbf{Conflict of Interest}}

The authors have no conflicts to disclose.

\ 
\

\noindent{\textbf{Data availability}}

The inputs and outputs related to the \gls{nep} model training are freely available at the Gitlab repository \url{https://gitlab.com/brucefan1983/nep-data}.

\end{document}